\documentclass[a4paper,11pt]{article}

\usepackage{graphics}
\usepackage{amssymb}
\usepackage{amsmath}
\usepackage{latexsym}
\usepackage{supertabular}
\usepackage{slashed}
\usepackage{booktabs}
\usepackage{cases}
\usepackage{bigints}
\usepackage{mysty}
\usepackage[abs]{overpic}

\newcommand{\nn}{\nonumber}
\newcommand{\simlt}{\lower.5ex\hbox{$\; \buildrel < \over \sim \;$}}
\newcommand{\simgt}{\lower.8ex\hbox{$\; \buildrel > \over \sim \;$}}
\newcommand{\resizeall}{\center\resizebox{0.8\textwidth}{!}}

\newcommand{\dsol}{\Delta_{21}}
\newcommand{\ddsol}{2\Delta_{21}}
\newcommand{\drct}{\Delta_{31}}

\newcommand{\datm}{\Delta_{32}}
\newcommand{\ssol}{\sin^22\theta_{12}}
\newcommand{\srct}{\sin^22\theta_{13}}

\newcommand{\dms}{\Delta m^2_{21}}
\newcommand{\dmr}{\Delta m^2_{31}}
\newcommand{\dcT}{\Delta\chi^2}
\newcommand{\dcTm}{(\Delta\chi^2)_{\rm min}}
\newcommand{\cT}{\chi^2}

\newcommand{\exposure}{${\rm 20 \,GW_{th}}$$\cdot$5kt (12\% free-proton
weight fraction)$\cdot$5yrs}
\preprint{KEK-TH/1581, KIAS-P/12067}
\title{Determination of mass hierarchy with medium baseline reactor
 neutrino experiments}

\renewcommand{\thefootnote}{\fnsymbol{footnote}}
\author[a]{Shao-Feng Ge\footnote{gesf02@gmail.com}}
\author[b,e]{Kaoru Hagiwara\footnote{kaoru.hagiwara@kek.jp}}
\author[c]{Naotoshi Okamura\footnote{nokamura@yamanashi.ac.jp}}
\author[d,e]{and Yoshitaro Takaesu\footnote{takaesu@kias.re.kr}}

\affiliation[a]{KEK Theory Center, Tsukuba, 305-0801, Japan}
\affiliation[b]{KEK Theory Center and Sokendai, Tsukuba, 305-0801, Japan}
\affiliation[c]{Faculty of Engineering, University of Yamanashi, Kofu,
 Yamanashi, 400-8511, Japan}
\affiliation[d]{Department of Physics and Astronomy, Seoul National
 University, Seoul 151-742, Korea}
\affiliation[e]{School of Physics, KIAS, Seoul 130-722, Korea}

\abstract{
We study the sensitivity of future medium baseline
reactor antineutrino experiments on the neutrino mass hierarchy. By using
 the standard $\chi^2$ analysis, we find that the sensitivity
depends strongly on the baseline length $L$ and the energy resolution $\left(\delta E/E\right)^2 =
\left(a/\sqrt{E/{\rm MeV}}\right)^2 + b^2$, where $a$ and $b$ parameterize the statistical
 and systematic uncertainties, respectively. The optimal length is found to be $L \sim
 40-55$ km, where a slightly shorter $L$ in the range is preferred for
 poorer energy resolution. The running time
 needed to determine the mass hierarchy also
depends strongly on the energy resolution; for a 5 kton
 detector (with 12\% weight fraction of free proton) placed at $L \sim 50$ km away
 from a $20 \,{\rm GW_{\rm th}}$ reactor, 3$\sigma$ determination needs 14 years of running with $a=3\%$ and $b=0.5\%$, 
which can be reduced to 5 years if $a=2\%$ and $b=0.5\%$. On the other
hand, the experiment can measure the mixing parameters accurately, achieving
$\delta \sin^22\theta_{12} \sim 4\times 10^{-3}, \delta (m_2^2-m_1^2) \sim 0.03\times 10^{-5} {\rm eV^2},$ and
$\delta |m_3^2 -m_1^2| \sim 0.007\times 10^{-3} {\rm eV^2}$, in 5 years,
almost independently of the energy resolution for $a < 3\%$ and $b < 1\%$.
In order to compare our simple $\dcTm$ results with those obtained by
simulating many experiments, we develop an efficient 
method to estimate the uncertainty of $\dcTm$, and the
probability for determining the right mass hierarchy by an experiment is
presented as a function of the mean $\dcTm$.  
}
\begin{document}

\maketitle

\renewcommand{\thefootnote}{\arabic{footnote}}
\setcounter{footnote}{0}
\renewcommand{\include}[1]{}
\renewcommand\documentclass[2][]{}

\documentclass[a4paper,11pt]{article}

\section{Introduction}
\label{intro}

Now that a large $\theta_{13}$ has been measured at Daya Bay~\cite{An:2012eh,*An:2012bu} and
RENO~\cite{Ahn:2012nd} experiments accurately, neutrino physics enters a
new era. One of the next challenges is determination of the mass
hierarchy. 
Many ideas have been proposed, such as 
long baseline accelerator-based neutrino oscillation~\cite{Minakata:2001qm,*Barger:2001yr,*Huber:2002mx,*Minakata:2003ca,*Blennow:2012gj,*Dusini:2012vc,
Chen:2001eq,*Aoki:2001rc,
Ishitsuka:2005qi,*Hagiwara:2005pe,*Hagiwara:2006vn,*Kajita:2006bt,*Hagiwara:2006nn,*Hagiwara:2009bb,*Hagiwara:2012mg},
atmospheric neutrino~\cite{PalomaresRuiz:2004tk,*Gandhi:2005wa,*Petcov:2005rv,*Blennow:2012gk}, 
supernova neutrino~\cite{Dighe:1999bi,*Minakata:2000rx,*Barger:2002px,*Lunardini:2003eh,*Dighe:2003be,*Dighe:2003jg,*Barger:2005it}, 
neutrino-less double-beta decay~\cite{Bilenky:1999wz,*KlapdorKleingrothaus:2000gr,*Bilenky:2001rz,*Pascoli:2001by,*Feruglio:2002af,*Pascoli:2002xq,*Pascoli:2002ae,*Petcov:2005yq,*Dueck:2011hu}, 
and medium baseline reactor antineutrino experiments~\cite{Petcov:2001sy,Choubey:2003qx,Learned:2006wy,Zhan:2008id,Batygov:2008ku,Zhan:2009rs,Ghoshal:2010wt}.

Among them, the medium baseline reactor antineutrino experiment has 
stimulated various re-evaluations of its physics potential and sensitivity
recently. 
Some works utilize the Fourier transform
technique~\cite{Ciuffoli:2012iz,Ciuffoli:2012bs,Qian:2012xh}, first
discussed in
refs.~\cite{Learned:2006wy,Batygov:2008ku,Zhan:2008id}, to
distinguish the mass hierarchy. The main advantage of this technique is
that the mass hierarchy can be determined without precise knowledge of the
reactor antineutrino spectrum, the absolute value of the large mass-squared difference $|\Delta m_{31}^2|$, 
and the energy scale of a detector. Although interesting and attractive, 
this technique is somewhat subtle to incorporate the uncertainties of
the mixing parameters and to estimate its sensitivity to
the mass hierarchy. On the other hand, some works adopt the $\chi^2$
analysis~\cite{Ghoshal:2010wt,Ghoshal:2012ju,Qian:2012xh} and new
measure based on Bayesian approach~\cite{Qian:2012zn}. These
methods utilize all available information from experiments, and it is
straightforward to incorporate the uncertainties to evaluate
the sensitivity, providing robust and complementary
results to the Fourier technique. 

In this paper, we analyze 
the sensitivity of medium baseline reactor antineutrino
experiments to the mass hierarchy for the baseline length of $10$--$100$ km and
the energy resolution $(\delta E/E)^2 =
\left(a/\sqrt{E/{\rm MeV}}\right)^2 +b^2$ in the range $2\% < a < 6\%$
and $b < 1\%$ with
the $\cT$ analysis. The optimal baseline length and the expected
statistical uncertainties of the neutrino parameters, $\ssol, \srct, \dms$
and $\dmr$, are also estimated. 

This paper is organized as follows. In Section \ref{basic}, we briefly discuss the
estimation of the energy distribution of reactor electron-antineutrino events at a far
detector. Section \ref{chi2} details the evaluation of the sensitivity
for determining the mass hierarchy using the $\chi^2$ analysis, and results of our analysis are shown in Section
\ref{sec:results}. In section \ref{fluctuation}, the statistical uncertainty of the sensitivity is
discussed, developing an efficient
method for estimating the uncertainty of the $\dcTm$. Finally, our
conclusions are summarized in Section \ref{conclusion}.
\bibliographystyle{bib2/general}

\bibliography{bib2/reference,bib2/neutrino}
\end{document}